\documentclass[sigconf]{acmart}

\AtBeginDocument{%
  }

\copyrightyear{2024}
\acmYear{2024}
\setcopyright{acmlicensed}\acmConference[DIS '24]{Designing Interactive Systems Conference}{July 1--5, 2024}{IT University of Copenhagen, Denmark}
\acmBooktitle{Designing Interactive Systems Conference (DIS '24), July 1--5, 2024, IT University of Copenhagen, Denmark}
\acmDOI{10.1145/3643834.3661591}
\acmISBN{979-8-4007-0583-0/24/07}

\usepackage{hyperref}
\usepackage{footnote} 
\usepackage{enumitem}
\usepackage{multirow}
\usepackage{tabularx}

\begin{document}

\title{PodReels: Human-AI Co-Creation of Video Podcast Teasers}

\author{Sitong Wang}
\affiliation{
  \institution{Columbia University}
  \city{New York}
  \state{NY}
  \country{USA}
}
\email{sw3504@columbia.edu}

\author{Zheng Ning}
\affiliation{
  \institution{University of Notre Dame}
  \city{Notre Dame}
  \state{IN}
  \country{USA}
}
\email{zning@nd.edu}

\author{Anh Truong}
\affiliation{
  \institution{Adobe Research}
  \city{San Francisco}
  \state{CA}
  \country{\normalsize USA}
}
\email{truong@adobe.com}

\author{Mira Dontcheva}
\affiliation{
  \institution{Adobe Research}
  \city{Seattle}
  \state{WA}
  \country{USA}
}
\email{mirad@adobe.com}

\author{Dingzeyu Li}
\affiliation{
  \institution{Adobe Research}
  \city{Seattle}
  \state{WA}
  \country{USA}
}
\email{dinli@adobe.com}

\author{Lydia B. Chilton}
\affiliation{
  \institution{Columbia University}
  \city{\normalsize New York}
  \state{\normalsize NY}
  \country{\normalsize USA}
}
\email{chilton@cs.columbia.edu}

\renewcommand{\shortauthors}{Wang, et al.}

\begin{abstract}
Video podcast teasers are short videos that can be shared on social media platforms to capture interest in full episodes of a video podcast. 
These teasers enable long-form podcasters to reach new audiences and gain more followers. 
However, creating a compelling teaser from an hour-long episode can be challenging. 
Selecting interesting clips requires significant mental effort; editing the chosen clips into a cohesive, well-produced teaser is time-consuming. 
To support the creation of video podcast teasers, we first investigated what makes a good teaser. 
We combined insights from audience comments and creator interviews to identify key ingredients. 
We also identified a common workflow used by creators during this process. Based on these findings, we developed a human-AI co-creative tool called PodReels to assist video podcasters in crafting teasers.
Our user study demonstrated that PodReels significantly reduces creators' mental demand and improves their efficiency in producing video podcast teasers.
\end{abstract}



\begin{teaserfigure}
\centering
\includegraphics[width=1\textwidth]{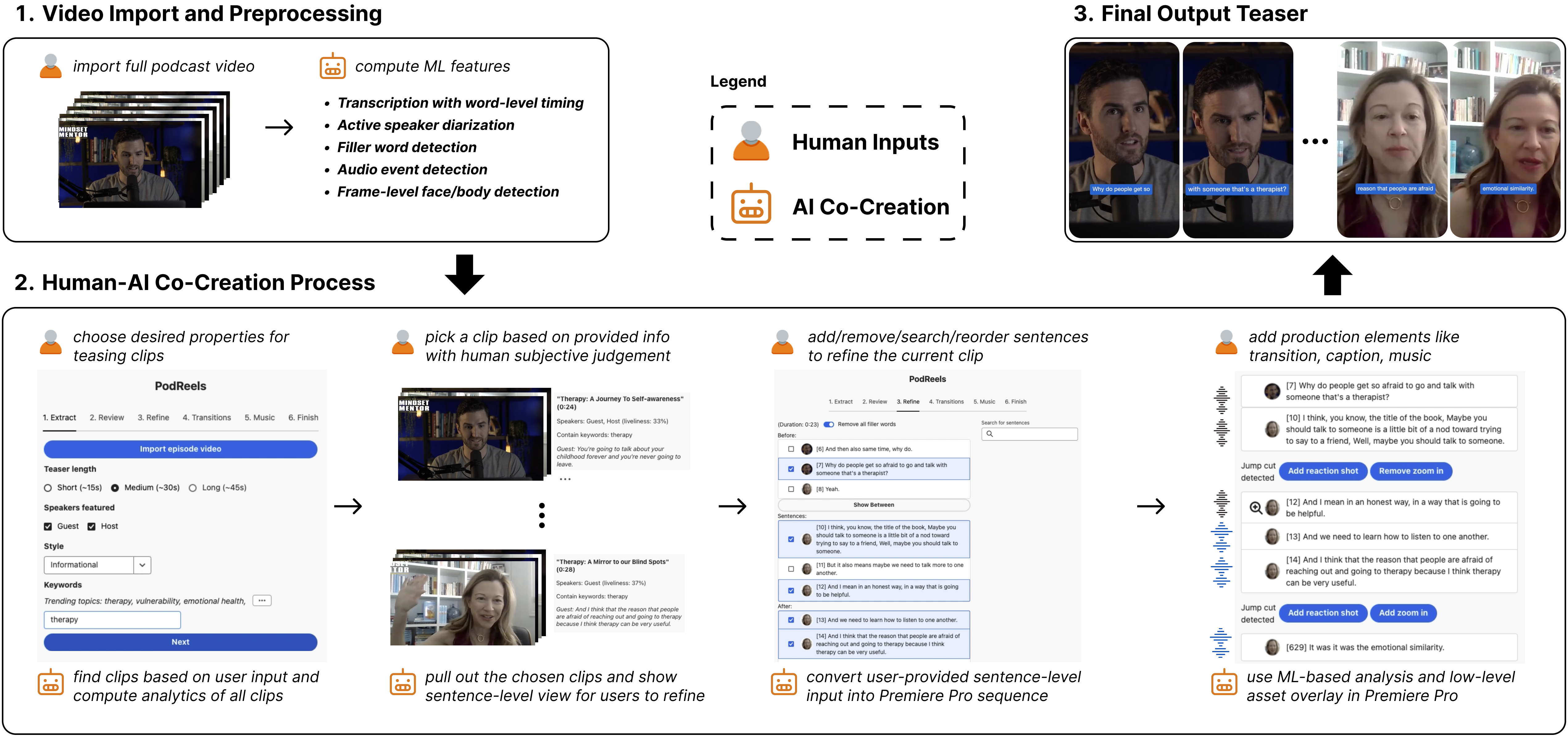}
\caption{
PodReels is a human-AI co-creative system that supports podcasters in creating video podcast teasers. 
PodReels takes the full-length long-form video as input and computes various ML features. 
The user provides desired properties of the teasing clips, including length, speakers to feature, style, and content keywords.
PodReels presents a list of candidate moments for the user. 
The user then fine-tunes the teaser structure and adds transitions, reaction shots, music, and captions to improve the production quality.
Finally, users can export the teaser and post it to social media platforms to promote their podcasts.}
\Description{PodReels is a human-AI co-creative system that supports podcasters in creating video podcast teasers. PodReels takes the full-length long-form video as input and computes various ML features. The user provides desired properties of the teasing clips, including length, speakers to feature, style, and content keywords. PodReels presents a list of candidate moments for the user. The user then fine-tunes the teaser structure and adds transitions, reaction shots, music, and captions to improve the production quality. Finally, users can export the teaser and post it to social media platforms to promote their podcasts.}
\label{fig:teaser}
\end{teaserfigure}

\maketitle

\section{Introduction }
Creating teasers is essential for content creators; 
seeing a short sample of content is a prime way to attract new viewers and continue to engage existing viewers. 
At the same time, creating teasers is a challenging design problem
because they must convey a ``representative emotional experience''~\cite{liu2018video} in a very short time. 
Most teasers are made through a content assembly process---where much or all of the content is selected clips from the source material. 
Picking the right clips from the source material requires balancing multiple design requirements: 
each clip must convey the right amount and kind of information, have the right emotional tone, and provide enough context so the user can understand the clip---all within 15 to 45 seconds.

Although creating movie trailers and teasers is an old problem, there is an emerging format of \textit{video podcasts}, and creators moving into this space need help making teasers for them.  
Video podcasts are filmed versions of podcasts. 
Hosts and guests film their conversations in the studio or with video conferencing software, and the videos are released to video platforms.
Traditionally, podcasts were an audio-only medium, but in recent years, the popularity of video podcasts has significantly increased.
For some podcasters, YouTube is now the most preferred platform for dissemination~\cite{Moreno2022}. 
According to a 2023 survey, 46\% of podcast listeners said they prefer consuming them with video~\cite{listerners_prefer_video}. 
A common way podcasters attract new audiences is through short teasers shared over video-based social media platforms such as TikTok, Instagram Reels, and YouTube Shorts.

To discover the challenges and approaches for creating compelling video podcast teasers, we conducted a formative study with five video podcast teaser creators and ten podcast listeners. 
We found creators often make short (around 30-second) teasers by selecting a single clip (or moment) from the source, then use music, cuts, transitions, and other effects to produce a teaser with high production value. 
Viewers value a high-quality teaser---both in content and production; 
for viewers to invest time in a podcast, they want a strong signal it will be worth their time.
For creators, selecting a clip from the source material is the biggest challenge: watching the podcast multiple times to identify clips is time-consuming, and balancing the trade-offs between different potential clips is mentally demanding. 
To achieve high-quality production value, they often add music, branding, and appropriate transitions. 
Creators want full control over the final output and need professional tools to polish the teaser.
To create a 30-second teaser for a 1-hour podcast episode, their overall process (including both content assembly and content production) takes about an hour. 
Reducing the time and mental demand would be valuable.

To assist the teaser creation process, we introduce a hybrid human-AI co-creation workflow embedded in professional video editing software.
Creators reported that there are many existing AI approaches for creating clips fully automatically, but they were generally dissatisfied with the results. 
These systems often produced an overwhelming amount of possibilities of generally low quality. 
Moreover, it was difficult to edit the results to add important context or remove rambling conversations. 
Instead, we envision AI to assist people in searching for clips, suggesting three clip windows based on multiple user-defined content criteria. 
Creators make the final selection of the clip as well as refining and editing the window. 
To evaluate this approach, we built PodReels, an extension to Premiere Pro that assists creators in rapidly selecting and refining interesting moments from their full-length episodes. 
PodReels combines automatic transcript analysis, speaker diarization, guided clip suggestions from large language models, and a simple editing interface to streamline the podcast teaser creation process (see Figure \ref{fig:teaser}). 

We evaluated PodReels in a within-subjects study with ten podcast creators who compared our system to a baseline interface that provided text-based editing but no clip suggestion, review, or refinement features.
Our work contributes:
\begin{itemize}[noitemsep,topsep=0pt]
    \item A formative study to understand the challenges and principles for creating engaging video podcast teasers.
    \item PodReels, a human-AI co-creative system that facilitates the content assembly and production process for video podcast teasers.
    \item A user study that shows using PodReels podcasters can create teasers in 59\% of the time with 44\% of the mental demand compared to a baseline system.
\end{itemize}

We conclude with a discussion on how human-AI tools within professional creative software can help achieve outputs of both better quantity and quality.
\section{Related Work}

\subsection{Movie Trailers and Teasers on Social Media}
Teasers have a long history in the movie industry. 
Movies use trailers to ignite anticipation and excitement for a new release among potential viewers.
A good trailer should provide a ``representative emotional experience'' and help the audience decide whether to watch the content~\cite{liu2018video}. 
Making movie trailers is a creative process that involves choosing shots based on aesthetics and story focus, arranging them in an engaging order without revealing spoilers, and selecting thematically consistent music~\cite{gaikwad2021plots}. 
Previous research has explored the task of movie trailer generation from the perspective of analyzing the movie's audio-visual features~\cite{smith2017harnessing}, maximizing visual attractiveness~\cite{xu2015trailer} and affective impact~\cite{irie2010automatic}.
Video podcast teasers focus less on visual aspects since they predominantly feature two or more people talking to each other, but they serve the same purpose of teasing the essence of the content and style, with the end goal being attracting a new and returning audience.

Teasers also play a significant role in transmitting informational content on social media platforms.
The task of teaser generation for disseminating news to social media users, specifically on platforms such as X (formerly Twitter), has been researched~\cite{news-teaser}. 
News teasers are bona fide, teasing, and abstractive, differentiating them from headlines (which are not teasing or abstractive) and clickbait (which is not bona fide)~\cite{news-teaser}. 
Tweetorial Hooks discusses the challenges of writing an engaging ``hook'' in the first tweet for a thread of tweets that explain STEM topics~\cite{tweetorials_cscw,tweetorial}. 

In this work we explore the task of creating teasers in the context of video podcasts. 
We learned the principles from related fields like movies,  news, and books~\cite{booktrailer101}, and conducted a comprehensive formative study to understand the creative strategies from the podcast audience and creators.

\subsection{Video Editing and Repurposing Tools}

Video editing is a complicated creative process involving manipulating media assets in the time and spatial dimensions. 
Compared to traditional timeline-based editing, transcript-based editing ~\cite{rubin2013content} makes editing videos with human speech easier, and as a result has been adopted by many professional video editing tools like Premiere Pro and DaVinci Resolve.
Recent research investigations use transcript-based methods to edit talking-head videos~\cite{fried2019text}, B-roll videos~\cite{huber2019b}, and dialog-driven scenes~\cite{roughcut}. 
Since music enhances content quality, much research has focused on music in video editing, for example, the use of musical underlays to emphasize key moments in spoken content~\cite{rubin2012underscore}, the manipulation of music to complement the speech~\cite{rubin2013content}, and the generation of emotional musical scores for audio stories~\cite{rubin2014generating}.

The video repurposing task involves editing and modifying an existing video to serve a different purpose or target a new audience and platform. 
To help audiences better digest information from videos, the video digest format was developed to aid in browsing and skimming informational videos~\cite{pavel2014}. 
Furthermore, an automatic multi-modal approach was used to generate hierarchical tutorials for navigating instructional videos~\cite{truong2021automatic}.
To cater to the creator's need to post their content on various platforms, the ROPE system~\cite{wang2022record} offers support by automatically trimming \emph{already short} audio stories (e.g., minutes long) at the sentence level. 
It creates even shorter versions that fit the length requirements of different social media platforms.

We are inspired by the transcript-based approach to video editing, in particular, the semantic understanding made possible by an accurate transcript. 
Taking it one step further, we focus on the creative task of repurposing hour-long video podcast episodes into short and compelling teaser reels. 
We summarized guidelines on what makes a good video podcast teaser and built an interactive system to co-create teasers with podcasters.

\subsection{Podcast Creation and Consumption Tools}
Podcasts have emerged as a popular medium for content creation and consumption. 
Podcast episodes, typically lasting for hours, are rich information sources. 
Much research has focused on the task of podcast summarization~\cite{li2021hierarchical, karlbom2020abstractive, song2022towards, rezapour2022makes}, which aids audiences in comprehending information without needing to listen to the entire episode. 
Although a short summary might seem similar to a teaser due to their similar lengths, they serve very different purposes.
From the audience's perspective, a good summary should cover a show's key points so well that they would skip it.
On the other hand, a good teaser needs to spark their interest enough that they feel compelled to watch the full episode.
Consequently, podcast creators are highly incentivized to make engaging teasers to increase the reach and growth of their podcast shows. 
In an interview with sixteen audio podcasters~\cite{rime2023will}, creators expressed a need for assistance in connecting with the audience, given the challenges of making their content discovered amidst a surge in podcast shows.

Video podcasts have become increasingly popular in the past few years~\cite{Moreno2022}.
Short video platforms, like TikTok, Instagram Reels, and YouTube Shorts, present new opportunities for podcast visibility. 
Many successful podcasts create teasers---short clips of podcast episodes---to draw in new audiences.
Start-up companies such as ClipsAI\footnote{\url{https://www.clipsai.com/}} and Opus\footnote{\url{https://www.opus.pro/}} are developing AI tools to help creators transform their long-form videos into short clips.
However, the output videos produced by these tools are often unsatisfactory and not up to the standards expected by experienced creators.
Additionally, the automatic generation approach employed by these tools leaves creators with very limited control over the process and the output. 
These automatic tools do not expose sufficient flexibility for creators to specify their needs and requirements, nor do they provide the capability to iterate over the output, leaving creators frustrated. 

PodReels supports podcasters to co-create teasers that match their creative input and gives creators flexible control over the creative process and results. 
We implemented the system as a Premiere Pro extension, leveraging existing video editing features (see Section \ref{sec:develop_plugin}).

\subsection{Human-AI Co-Creativity}
Generative AI has shifted the landscape of creativity support tools. 
Large language models (LLMs) like GPT-4~\cite{gpt}, which have been trained on a vast corpus of documents and embody broad general knowledge, are powerful in developing and prototyping many design ideas quickly.
Recent creativity support tools have used generative AI in supporting various content creation tasks, including writing~\cite{sparks,talebrush,anglekindling}, image generation~\cite{opal, popblends}, and video creation~\cite{disco,reelframer,lave, moodsmith}.

Though powerful, these LLMs are not perfect.  GPT-4 is known to hallucinate and provide false information~\cite{gpt}. 
Also, generative AI cannot replace creators' subjective evaluations of the content it creates.
Therefore, it is important to develop human-AI co-creation systems that bring human judgments, norms, and values into the creative process. 
Many human-AI co-creation systems facilitate the creation process by streamlining workflows that involve the user at each step. 
For example, to support argumentative writing, VISAR~\cite{zhang2023visar} implements visual planning and utilizes LLMs to generate key aspects, discussion points, and argumentative sparks for outline nodes. 
To aid in creating illustrations for a headline, Opal~\cite{opal} first employs LLMs to help users explore keywords, tones, and artistic styles, and then generates images using the determined subject and style with DALL-E.
To support journalists in translating news articles into social media reels, ReelFramer~\cite{reelframer} provides multiple narrative framings for users to explore and helps users establish the characters, plot, setting, and information points before creating the reel script.
At each step, AI provides suggestions, but users can always regenerate, edit, or write their own content to maintain control over the process.
In the human-AI co-creation process, AI can brainstorm and make prototypes quickly, while humans need to be involved in each step to guide the process and edit to ensure the content is coherent and appropriate.

We introduce a human-AI co-creation experience to combine the intelligence of humans and AI. 
We utilize generative AI and other machine learning algorithms to ease the creator's burden of searching and prototyping, while involving humans in each stage for evaluation and guidance. 
We use LLMs for \emph{extractive} summaries, sidestepping the issue of hallucinating non-existent material from \emph{abstractive} summaries.
Additionally, we embed this tool directly into professional video editing software to give creators the option for full control whenever needed. 
\section{Formative study}

To understand the ingredients of successful video podcast teasers and the creation process, we conducted interviews with both podcast listeners and creators.
We mainly investigated three questions:
\begin{enumerate}
 \item What makes a good video podcast teaser? 
 \item What is the current workflow of creating video podcast teasers? 
 \item Why is it hard to create a good video podcast teaser?
\end{enumerate} 

\emph{Listeners and teaser examples.} We invited ten podcast listeners (L1-10, average age 23.4, five female, five male) through word-of-mouth.  
 All listeners were video podcast enthusiasts who watched podcasts daily.
 
 We collected seven video podcast teaser examples (see Table \ref{tab:exsiting_examples}). 
We checked popular podcast channels on YouTube\footnote{\url{https://www.youtube.com/podcasts/popularshows}}, manually verifying which have teasers and randomly selecting an episode from each channel.
We asked participants to watch the examples and provide qualitative feedback on their likes and dislikes for each teaser. 
We also checked whether the teaser enticed them to watch the entire episode. We paid each participant \$25/hour.

\emph{Creators.} We recruited five expert video podcast teaser creators (C1-5, average age 27.6, three female, two male, background see Table \ref{tab:formative_interviewees}).
We recruited them by posting on social media, such as Facebook groups for podcasters, and on Upwork\footnote{\url{https://www.upwork.com/}}.
On average, each creator had approximately two years of experience creating teasers for video podcasts across various genres. 
Each had produced at least 50 video podcast teasers. 
During the interviews, they were asked to showcase their work and share their strategies for creating the teasers. 
They also walked us through their creation process and typical workflow.
We paid each participant \$50/hour.

We applied the thematic analysis method~\cite{braun2006using} to analyze the interview transcripts. 
The initial coding was performed by the first author through several iterative rounds to identify emerging topics from the interviews.
Subsequently, the research team collaboratively reviewed the coding results to determine the themes.
We present the key findings as follows.

\begin{table*}[t]
    \begin{tabular}{| c | c | c | c | c | c | c | }
    \hline 
         &  \textbf{Podcast show} &  \textbf{Podcast genre} & \textbf{Episode title with teaser link} &\textbf{Single/Multi-moment} &\textbf{Teaser length}\\ 
         \hline
         \textbf{E1}&  Impact Theory &Education& \href{https://www.youtube.com/watch?v=La9oLLoI5Rc}{How to Brainwash Yourself...} & Single-moment &  30s \\ 
         \hline 
          \textbf{E2}& Decoder & Business& \href{https://www.youtube.com/watch?v=cK3nR77RBsU}{GaryVee vs. the Real Gary...} & Multi-moment  & 1min 54s\\ 
         \hline
         \textbf{E3}&  Lex Fridman Podcast&Technology& \href{https://www.youtube.com/watch?v=sY8aFSY2zv4}{Jordan Peterson: Life, Death...} & Single-moment & 22s \\ 
         \hline 
        \textbf{E4}&  Jay Shetty Podcast &Health\&Fitness& \href{https://www.instagram.com/reel/Csdwo-epsxp}{Kim Kardashian Opens Up...} & Multi-moment &  1min 7s\\ 
         \hline 
         \textbf{E5}&  Your Mom's House&Comedy& \href{https://www.instagram.com/reel/Ct4fMtHJI2E}{Exhibitionist w/ Lauren...} & Multi-moment & 56s\\ 
         \hline 
         \textbf{E6}&  Off the Vine&Society\&Culture& \href{https://www.instagram.com/reel/Ctt-Q_6u8uI}{Charity Lawson: Bachelorette...}   & Multi-moment & 55s\\ 
         \hline 
         \textbf{E7}&  Asian Boss Girl&Society\&Culture& \href{https://www.youtube.com/watch?v=EG5ppWwSUgk}{Our Work Besties and Janet's...} & Single-moment & 20s\\ 
         \hline 

    \end{tabular}
    \caption{Examples of existing video podcast teasers collected from popular podcast channels. The genre information was collected from the description of the corresponding podcast channel on Apple Podcasts.}
    \Description{Table showcasing examples of video podcast teasers from various popular podcast channels, including details such as Podcast show, Podcast genre, Episode title with teaser link, whether the teaser is a Single or Multi-moment, and the Teaser length. Genres range from Education, Business, Technology, Health & Fitness, Comedy, to Society & Culture. Each row represents a different podcast episode, with links to teasers, and specifies if those teasers capture a single moment or multiple moments from the episode, along with the duration of each teaser.}
    \label{tab:exsiting_examples}
\end{table*}

\begin{table*}[t]
\begin{tabular}{| c | c | c | c |}
\hline
 & \textbf{Creator role}&\textbf{Podcast genre} &\textbf{Video teaser format}\\
 \hline
 \textbf{C1}&  Studio producer and editor& Comedy &Instagram\\
 \hline
 \textbf{C2}&  Content producer& News &Pre-pended video, Instagram, TikTok\\
 \hline
 \textbf{C3}&  Host, producer and editor& Society\&Culture &Pre-pended video, Instagram\\
 \hline
 \textbf{C4}&  Freelance teaser creator& Business &YouTube Shorts, Instagram, Facebook\\
 \hline
 \textbf{C5}&  Freelance teaser creator& TV\&Film &YouTube Shorts, Instagram \\
\hline
\end{tabular}
\caption{Background information of the expert creators. On average, each creator had about two years of experience creating video podcast teasers. Each had produced at least 50 video podcast teasers.}
\Description{Table summarizing expert creators' backgrounds in producing video podcast teasers. It consists of four columns: 'Creator role,' 'Podcast genre,' 'Video teaser format,' and a row identifier ranging from C1 to C5. Each creator is associated with a specific podcast genre, such as Comedy, News, Society & Culture, Business, and TV & Film, and uses different video teaser formats including Instagram, TikTok, YouTube shorts, and Facebook. Their roles vary from studio producers and editors to content producers, hosts, and freelance teaser creators. On average, each has two years of experience and has produced at least 50 video podcast teasers.}
\label{tab:formative_interviewees}
\end{table*}

\subsection{What makes a good video podcast teaser?}
We identified four key principles for creating a good video podcast teaser, based on interviews with both the audience and creators. 

\subsubsection{Short duration}
\label{sec:duration}
%[Take away: 15-45 seconds target length; single-moment teaser]
Short duration is an important element for video podcast teasers. 
A recurring theme from interviews with creators underscores the need for brevity and conciseness.
For example, C2, who has extensive broadcasting experience, shared that the structure of newscast teasers has remained constant over the years---an effective teaser should be brief, around 30 seconds, and captivating. 
Similarly, C5 noted that at the start of their career, they made teasers that were too long and did not yield good results. 
However, they observed a significant improvement when they made their teasers concise and to the point.
One strategy that all creators we interviewed adopt is creating single-moment teasers, which highlight one single compelling moment from the episode, ensuring the teaser remains short and focused.

Listeners felt very similar to creators. They wanted short, focused teasers. 
For example, eight out of ten listeners thought the teaser for E2 was too long, lasting 1 minute and 54 seconds. 
In addition, listeners preferred single-moment teasers (E1, E3, E7), with a common sentiment of being ``short and to the point'' and ``efficient in teasing the episode''.
To understand the range of an efficient duration, we collected 30 additional examples from the three popular podcast channels that create single-moment teasers (as shown in Table \ref{tab:exsiting_examples}), each channel contributing ten examples.
We found the teaser duration range is 11 to 51 seconds, with an average of 27.2 seconds (SD=13.1).

\textit{Therefore, a good video podcast teaser should have a short duration of around 30 seconds and be focused. 
An efficient strategy is to highlight a single moment from the episode.} 

\subsubsection{Strong hook}
\label{sec:hook}
%[Take away: hook should be attention-grabbing, tease the podcast style and topic, and also authentic]
A strong hook is another key element in creating compelling teasers. 
In the context of video podcast teasers, a strong hook is a captivating snippet that quickly grabs the audience's attention and compels them to explore the whole episode. 

The quality of the hook in a teaser often determines whether a potential listener will continue scrolling through their social media feed or click to watch.
Both C4 and C5 mentioned that they would put themselves in the shoes of a new listener to choose the most appealing snippet. 
This snippet should catch their attention and prevent them from scrolling past on their phone, keeping them interested. 
The chosen hooks should align with the podcast's genre, as the creators emphasized during their interviews. 
For instance, C1, a producer and editor of a comedy podcast, highlighted the importance of keeping the teaser funny. 
While C3, a society and culture podcast host, often features the most touching parts of the episode in their teasers.

The importance of strong hooks is also echoed in listener analysis of teaser examples. 
According to listener's feedback, a powerful hook can quickly seize their attention with its engaging style. 
Several strands of style-specific keywords were frequently mentioned in audience comments about the hooks they found appealing. 
For podcast episodes that primarily serve information needs (E1, E2, E3), the audience described the hooks they liked as ``informational'' (similar keywords include ``insightful'', ``thought-provoking'') or ``curiosity-arousing'' (similar keywords include ``intriguing'', ``counterintuitive''). 
While for episodes that cater more towards entertainment (E4, E5, E6, E7), the audience said they liked the hooks because they were ``funny'' (similar keywords include ``cheerful'', ``humorous'', ``amusing'') or ``emotional'' (similar keywords include ``touching'', ``sensitive'').
Additionally, the authenticity of a hook is critical. 
It should not only captivate but also accurately reflect the substance of the podcast content. 
A hook that feels very ``clickbaity''  may have a counterproductive result, potentially eroding trust in the podcast brand.

\textit{Therefore, a good video podcast teaser should effectively utilize a strong hook that is both attention-grabbing and authentic, aligning with the podcast's genre and style. Effective styles include informational, curiosity-arousing, funny, and emotional.}

\subsubsection{Clear context}
\label{sec:clarity}
%[Take away: teaser should be self-contained, straightforward, give basic background, not confusing]
Providing clear context is crucial when creating self-contained video podcast teasers. 
All creators shared that a good teaser should set the stage for the audience, providing them with enough information to understand what the podcast or a specific episode is about. 
This helps guide audience expectations and clarifies the content's relevance or value.
However, providing clear context does not mean revealing all episode details in the teaser.

The key is to strike a balance: reveal enough to attract and engage the audience, but withhold enough details to arouse curiosity and anticipation. 
A counter-example is E6. 
According to the audience, this teaser was ``very unclear about the topic and if the guest really answered the questions''(L8) and ``has no takeaway at all''(L7). 
The lack of clear context makes half the audience decide they would not spend time checking the whole episode.

In addition, the content should be easy to understand for a listener who is brand new. 
C2 stated that the podcast episode typically consists of an hour-long dialogue. 
Naturally, in the midst of such a conversation, the speaker may not express everything clearly. 
For instance, it is common for speakers to use referents like ``here'' and ``them''.
A teaser with many of these out-of-context words makes it difficult for a new viewer to understand what is happening, thereby failing to communicate the message effectively.
For example, eight out of ten audiences mentioned feeling confused when watching E7 (the remaining two happened to have watched the show before). 
This teaser includes many out-of-context conversation pieces from the middle of the episode, providing the audience with little useful information about the speakers and hindering their ability to decide whether to check out the full episode.

\textit{Therefore, a good video podcast teaser should reveal enough information to attract and engage the audience, and withhold enough details to arouse curiosity and anticipation.
To be easily understood by new viewers, a good teaser avoids out-of-context content that may confuse them.}

\subsubsection{High production quality}
%[Take away: production is necessary and beneficial]
All creators emphasize the importance of production, viewing it as an essential aspect of adhering to target platform styles and highlighting the podcast brand. 
Furthermore, production serves as another opportunity to capture the audience's attention. 
For example, as C5 pointed out, the visuals of podcast videos are usually static---investing effort into production can make the teaser more visually engaging.
All creators we interviewed used professional editing tools for production.

Production quality is also a key factor that all viewers mentioned they would consider when deciding whether or not to watch the podcast. 
Even if the specific topic or guests in the episode do not interest the audience, they may still explore the podcast channel if the teaser shows high production quality. 
Listeners widely appreciate the visual dynamics created by elements such as shifting focus to the current speaker and using motion graphics in captions. 
The audience also mentioned that the appropriate integration of background music significantly enhances their focus on the content. 
Unique branding designs, like preset openings or logos, also garner the audience's respect, as seven out of ten participants stated.

A teaser example of stellar production quality is E4.
This teaser incorporates multiple audio elements like sound effects and music, visual components like motion graphics, B-rolls, reaction shots, and captions highlighting key phrases or words. 
All of these are edited with seamless transitions. 

\textit{Therefore, a good video podcast teaser should have high production quality, adapt to the platform style, ensure a smooth flow in audio and visual dimensions, and appropriately use other assets to boost engagement.}

\subsection {What is the current workflow for creating video podcast teasers? }
Creators shared the tools, strategies, and steps they currently use to create video podcast teasers.

\subsubsection{All interviewed creators use professional non-linear video editing tools}
All creators we interviewed use professional non-linear video editing tools for their creation process, including Adobe Premiere Pro and Final Cut Pro. 
Thus, it makes sense to develop support for the teaser creation process within \textit{professional, non-linear video editing tools}, where creators are already accustomed to working.

\subsubsection{All interviewed creators use a single-moment style} 
\label{sec:creator-single-moment}
All five creators make teasers using a single-moment style. They find this approach an efficient way to create short and focused teasers, as discussed in Section \ref{sec:duration}.
Although one creator (C4) expressed interest in trying multi-moment teasers in the future, the other four disagreed. 
They considered creating these multi-moment teasers labor-intensive for a weekly-release podcast show (C1). 
More importantly, they felt the teasers tended to condense too much information into one single clip (C3), were often too long to watch (C5), and diluted the teaser's effectiveness (C2).
Thus, supporting the creation of \textit{single-moment teasers} is optimal due to their efficiency and focus. 
Still, the mechanism of creating multi-moment teasers could be developed based on the single-moment structure for creators who wish to experiment.

\subsubsection{All interviewed creators use a common workflow for content assembly and production}
We identified a common workflow amongst all five creators. 
In the content assembly phase, creators \textit{extract} multiple candidate moments, either by pulling sections from the editing software during post-edits or by noting timestamps for later reference. 
They then \textit{review} these moments to make a second round of selection, opting for one moment.
Subsequently, they \textit{refine} the chosen clip for the desired length, pacing, and clarity of context. 
In the production phase, all creators add \textit{transitions} to ensure smooth flow, most (four out of five) add \textit{music} to enhance engagement, and all \textit{finish} by adjusting the aspect ratio and adding captions and logos. 
Depending on the episodes, several other production steps might also be taken, like adding sound effects and B-rolls.
Thus, there are two phases in the workflow: content assembly, where users \textit{extract, review, and refine}; and production, where users add \textit{transitions, music, and finish} the teaser.

\subsubsection{Strategies for content assembly}
\label{sec:strategies}
Creators agreed that the most difficult and time-consuming part of the process was content assembly.
In the assembly step of their workflow, creators consider \textit{duration} (Section \ref{sec:duration}), engagement (Section \ref{sec:hook}), and clarity (Section \ref{sec:clarity}). 
For engagement, all creators shared that they would pick the hook moments of an engaging \textit{style} that fits the genre of their episode.
Three creators also shared that they would choose ``topical'' moments related to a trending topic---a common strategy is to find snippets that cover the topic \textit{keywords} of the episode.
For clarity, three creators select moments that cover identifiable keywords, making the subject of discussion clear to new listeners.
In addition, all creators consider the \textit{speakers featured} in the teaser.
C2 prefers to highlight hosts more, viewing them as the brand of their news podcast. 
While C3, C4, and C5 deem it crucial to give prominence to the guests, C1 aims to cover both hosts and guests.
Thus, we identify \textit{duration}, \textit{style}, \textit{keywords}, and \textit{speakers featured} as the four essential dimensions to support in identifying compelling moments.

After selecting potential moments, all creators review the candidates' audio quality. 
C2 pointed out that, in an hour-long episode, there will be times when the speakers sound tired and lack energy. 
Creators prioritize more ``lively'' portions of a conversation, avoiding parts where the individual's voice comes across as monotonous. 
Thus, we identify \textit{``liveliness''} as another key factor that creators consider in the review step.

After selection comes refinement. C3 points out that they rarely can directly use a single portion of the episode as the teaser.
They need to bring in neighboring segments to make the context clearer. 
For example, even if the picked portion is engaging in terms of style, creators might still want to incorporate a few more surrounding sentences to ``make the main topic more evident'' (C2). 
Also, since keeping the short duration is important for a successful teaser, all creators mentioned spending time removing dead parts of the conversation, such as filler words and sentences where the speaker rambled a lot, and adjusting the order to make sure the teaser is focused and flows at a fast pace.
Thus, we identify \textit{finding surrounding statements, adjusting the clip at the sentence level, and removing filler words} the primary actions users took in the refining step.

\subsubsection{Those who have tried existing AI tools for teasers want more control and transparency.}
\label{aitool_opinion}
Two creators, C4 and C5, shared their experiences using AI automated tools, including Clips.AI and Opus, to aid their creative processes.
Although they saw potential in using AI to support them in creating teasers, they did not incorporate these tools into their workflows.  
For example, C5 found that Clips.AI produced too many results, overwhelming the review process. 
C5 described how an attempt to use the tool on a two-hour podcast episode yielded almost 80 results, making it just as inefficient as watching the entire episode. 
While Clips.AI ranks the results based on what it perceives as the most compelling moments, C5 felt the accuracy was questionable. 
The reasons given by the tool were often vague, which confuses the review and decision-making processes.

Opus was more successful in not overwhelming the creators by restricting the output to 15 clips, each paired with a score.
However, C4 found the quality of the outputs to be inconsistent.
Opus also automatically edited the moments, which was not well-received by C4. 
To them, the edited clips appeared amateurish, basic, and generic.
The clips all seemed identical and lacked creativity, undermining what C4 regards as the essential purpose of content creation: ``creative expression''.
C4 would prefer if Opus simply suggested the raw selected content instead of edited clips, giving them greater control over the production.
Hence, an ideal AI-powered creativity support tool should \textit{select multiple, but not an overwhelming number of clip options, and help creators understand why they are chosen}.
The tools should also not be entirely automated and instead should \textit{involve users and provide them with control over the process and results}.

\subsection{Why is it hard to create a video podcast teaser?}
 
Creators also shared the pain points of the process, identifying the steps they found most cognitively demanding or time-consuming.

\subsubsection{Picking compelling moments is cognitively challenging}
All participants consider \emph{picking moments} the most cognitively challenging step.
Even experienced creators find it hard to locate moments that fit all the desired parameters.
They often need to rewind and review the hour-long episode a lot to identify the compelling moment candidates.

\subsubsection{Pulling out and refining segments is time-consuming}
\label{sec:pullout-time}
After picking some moments, creators need to pull out these segments. 
Three creators reported this as the most time-consuming step, which involves recording timestamps while (re)watching the episode and locating them later.
Timeline controls often lack easy controls for extracting a short segment from a long video. 
Users must zoom in and out on the timeline to find and extract each segment.

Four out of five creators mentioned that it takes time to find and combine segments to refine the moment. 
Searching for additional segments for refinement is time-consuming as creators need to replay and pinpoint the exact locations. 
Taking duration and flow into consideration, they also need to identify and remove sentences where speakers are rambling. 
All these contribute to making manual teaser creation a laborious process.

\subsubsection{Production steps take lots of effort}
Creators spend a significant amount of time on productions. 
All participants mentioned adjusting the aspect ratio and adding captions to fit the targeted social media platforms. 
They also removed abrupt jump cuts, long silences, or filler words to create a smooth flow. 
In addition, four out of five creators added extra assets, such as background music and channel logos, to enhance engagement and branding. 
All these production steps require intensive manual work on a timeline. 
Creators must collect extra assets, navigate through the timeline, and carefully cut and assemble all materials in audio and visual dimensions cohesively.

\subsection{Design Goals}
In conclusion, we identify four design goals our tool should support:

\begin{enumerate}[nosep,leftmargin=*,label={\textit{D{\arabic*}}}]
\item \textit{Support content assembly and production of single-moment teasers}: 
Single-moment teasers are valuable to support, as they are short, to the point, and preferred by most creators interviewed.
Our tool should streamline common steps in content assembly and production for creators.

\item \textit{Support the search for hooks that meet multiple design criteria}: 
Our tool should allow for multi-dimensional searches, cover the design parameters mentioned by creators, and provide multiple options in a non-overwhelming manner.

\item \textit{Ease the process of refining and pulling out segments with context}: 
Our tool should allow for transcript-based editing and offer a flexible ``window''-based adjustment for users to refine the context.

\item \textit{Support production of all essential elements and give users control}: 
Our tool should help creators within the professional editing tools they are already using, offering support for content assembly and production, while also giving them the control to go further if needed.
\end{enumerate}

\section{PodReels System}

PodReels is a human-AI co-creation system that supports podcasters in creating video podcast teasers. 
The system leverages AI to efficiently search and prototype ideas, while humans are engaged throughout the process to edit, evaluate, and make high-level decisions.
It adopts a text-based editing method because podcast content is primarily spoken words and editing words is much easier to do with text than a traditional non-semantic video timeline.
PodReels uses sentences as the basic editing unit because sentences have a natural starting and ending point of speech and often express a complete idea/sub-idea.
Within the paradigm of text-based video editing,
we facilitate human-AI co-creation by using several AI methods to produce high-level operations that creators need for content assembly and production. 

PodReels is powered by a machine learning feature ingestion pipeline and a large language model, GPT-4 
(gpt-4-32k-0613 model\footnote{\url{https://platform.openai.com/docs/models/gpt-4-turbo-and-gpt-4}}). 
At the time of our deployment, we chose this model because its 32k-token context window can accommodate the length of a multi-hour podcast transcript.
We apply zero-shot prompting in the system.
The prompts were initiated by the first author and then tested and iterated with the research team before being finalized.

The ingestion pipeline processes videos from both the auditory and visual modality and includes the following ML algorithms: transcription with word-level timing, active speaker diarization~\cite{alcazar2020active}, which assigns each sentence to a speaker, filler word detection~\cite{zhu22e_interspeech}, audio event detection~\cite{serizel2020sound}, and frame-level face/body detection~\cite{varol2018bodynet} (see Appendix for the mapping of these technical components to the features of PodReels).

To support content assembly and production of single-moment teasers (D1),
PodReels streamlines the workflow of six main steps 
1) Extract multiple interesting moments, 
2) Review the moment candidates and pick one to continue with, 
3) Refine the selected moment to enhance clarity and engagement, 
4) Add transitions to hide jump cuts, 
5) Add music that aligns with the speech content, 
and 6) Finish with production polishing steps such as adjusting the aspect ratio, adding captions and a logo page.

\subsection{Extract}
\label{sec:extract}

When a user imports an input podcast video, the system loads the transcription with speaker diarization data and creates a new sequence for the imported video. 
To support the search of hooks that meet multiple design criteria (D2), users specify the following elements that the system will try to accommodate, as shown in Figure~\ref{fig:teaser} (bottom left):

\begin{itemize}
\item\textbf{Desired length} 
The default length is 30 seconds, as discussed in Section~\ref{sec:duration}. 
We also support lengths of 15 seconds and 45 seconds for flexibility.
\item\textbf{Featured speakers}
Typically, the guest(s) is featured more prominently in the teaser. However, we also allow the user to select to feature the host(s) or both the guest(s) and the host(s).
\item\textbf{Style} The user can choose the style of the clip to be informational, curiosity-arousing, funny, or emotional, depending on the episode's genre and theme.  
\item\textbf{Keywords} To highlight specific content, users can add relevant keywords. 
The system provides topic keyword suggestions for reference. 
These keyword suggestions are derived from the episode content by the LLM and ranked based on the frequency of Google Trends searches over the past week. 
Google Trends is a reliable trend indicator of the popularity of search terms and topics, making it feasible to evaluate a topic's trending value.
\end{itemize}

To extract moments PodReels constructs a prompt and feeds the LLM the video transcript (see the prompt in Appendix).
The LLM returns a list of sentence IDs corresponding to the clips it picks.
The system then maps back these sentence IDs to video timestamps.

PodReels interfaces with Premiere Pro APIs to trim and generate clips corresponding to the sentences. 
All clips are placed on the timeline automatically, however this interaction between PodReels and Premiere Pro happens 100\% in the background. 
Users can look at the timeline but they do not need to.
From the user's perspective, they only need to decide which moments to use and proceed by clicking ``Next''. PodReels presents the moment suggestions.

\begin{figure}
\includegraphics[width=0.47\textwidth]{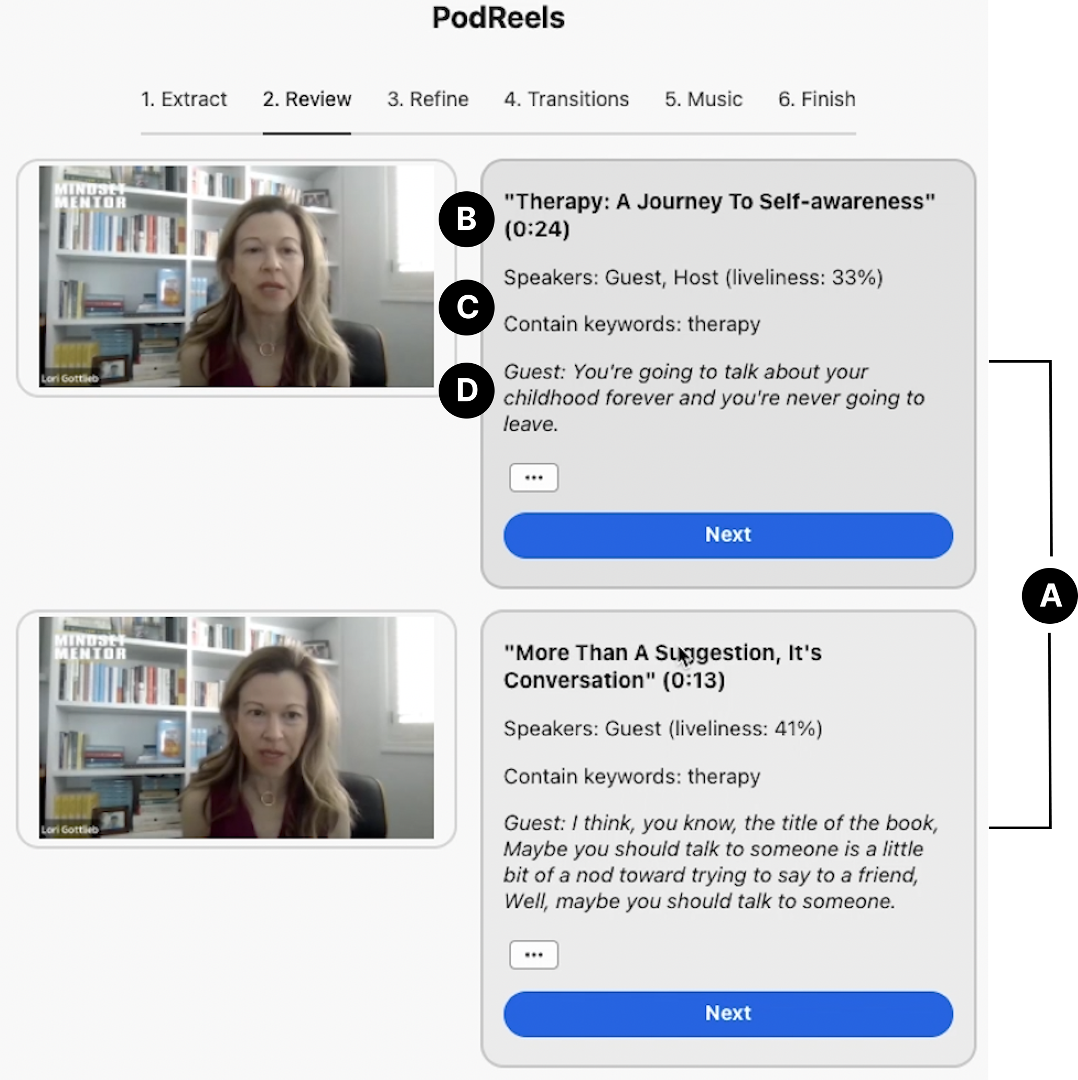}
\caption{The review step helps creators pick the right moment for their teaser. (A) Based on user-provided queries (e.g., desired length), we show a list of candidate moments for users to scroll and review. For each moment, we provide a compact yet informative view to help users make a decision. (B) We show an AI-generated tagline for each moment and include a duration. (C) We also show who the active speakers are, along with a liveliness score computed from audio features. We also verify whether the input keywords are present in the moment. (D) Finally, we show an expandable full transcript view, allowing users to glance through.}
\label{fig:review}
\Description{The review step helps creators pick the right moment for their teaser. (A) Based on user-provided queries (e.g., desired length, keywords, style, etc.), we show a list of candidate moments for users to scroll and review. For each moment, we provide a compact yet informative view to help users make a decision. (B) We show an AI-generated tagline for each moment and include a duration. (C) We also show who the active speakers are, along with a liveliness score computed from audio features. We also verify whether the input keywords are present in the moment. (D) Finally, we show an expandable full transcript view, allowing users to glance through.}
\end{figure}

\subsection{Review}
\label{sec:review}

The system provides a list of clip options for users to review and pick from (see Figure \ref{fig:review}).
For each moment, the system generates a moment overview to support efficient review.
Upon clicking the thumbnail in the overview, PodReels plays the moment.
To aid the user in swiftly scanning and assessing the quality of the suggestions, PodReels includes the following information in each overview (D3):

\begin{itemize}
\item \textbf{Clip tagline}. The system provides users with a clip tagline, to help them quickly grasp the main theme of the clip. The tagline is generated by feeding the clip transcript text to the LLM (see the prompt in Appendix).
\item \textbf{Duration}. The system calculates and displays the clip's duration using word-level timing data.
\item \textbf{Speakers and audio quality}. The system indicates which speakers were included in the moment based on the speaker diarization data. 
In a short teaser, creators would want to feature the most vibrant conversation. 
We compute their speaking liveliness by averaging the audio amplitude of the clip~\cite{wang2018effect}.
\item \textbf{Keywords}. The system indicates whether the selected clip includes the keywords entered by the user. 
The system searches the transcript for the candidates and displays the contained keywords, helping the user in determining if their search criteria have been met.
\item \textbf{Transcript preview}. The system also provides a transcript preview to facilitate the user in rapidly scanning the actual content of the moment. 
By clicking the ``...'' button, they can view the full transcript of the clip.
\end{itemize}

The system displays three candidate moments for the user to review. 
If they wish to view additional suggestions, they can click the ``Show More'' button.
If they like a suggestion, they can click the ``Next'' button to continue. 
If none of the clips matches their goal, they can return to the Extract tab to adjust the extraction parameters, e.g., trying a different keyword or style.

\subsection{Refine}

\begin{figure}
\includegraphics[width=0.48\textwidth]{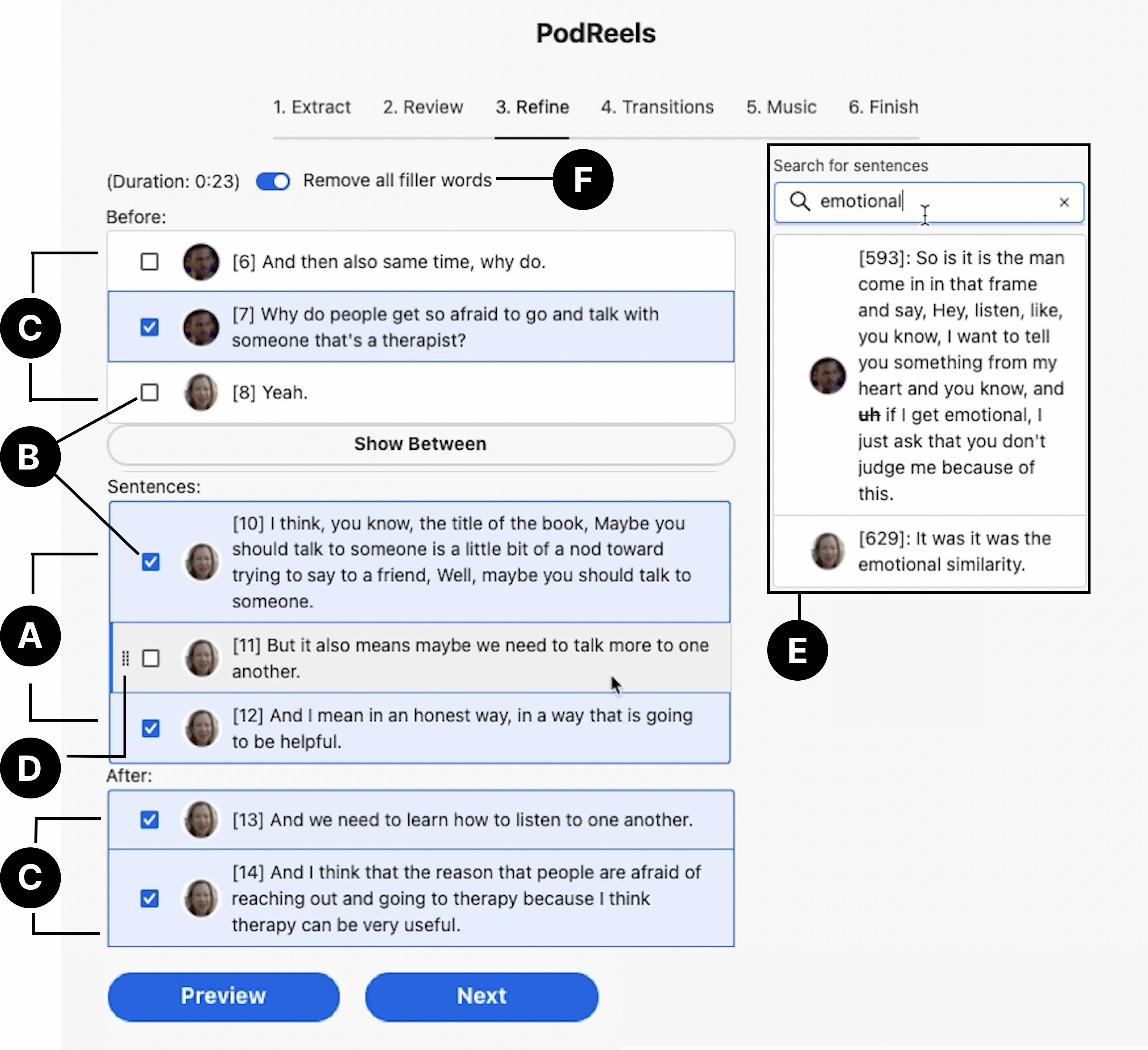}
\caption{In the refine step, we show a sentence-based view. (A) By default, the center portion, which represents the sentences within the selected moment, are selected. (B) Users can toggle on or off each sentence to include or exclude from the teaser. (C) To provide better context, users can pick from the before and after surrounding sentences and add them to the teaser. (D) Users can reorder the sentences to improve the flow. (E) If users want to add a sentence that is not listed, they can use the search box to find the sentence from the full transcript and drop it to insert. (F) Users can also remove all filler words with a toggle switch.}
\label{fig:refine}
\Description{In the refine step, we show a sentence-based view. (A) By default, the center portion, which represents the sentences within the selected moment, are selected. (B) Users can toggle on or off each sentence to include or exclude from the teaser. (C) To provide better context, users can pick from the before and after surrounding sentences and add them to the teaser. (D) Users can reorder the sentences to improve the flow. (E) If users want to add a sentence that is not listed, they can use the search box to find the sentence from the full transcript and drop it to insert. (F) Users can also remove all filler words with a toggle switch.}
\end{figure}

To ease the process of refining and pulling out segments with context (D3), we provide transcript-based video editing and a flexible "window”-based adjustment for users to refine the context.
Any individual moment will likely need refinement to be appreciated out of context. 
It might have sentences not vibrant enough which the user wants to remove. 
It might have ended a bit too early, and the user would like to add additional context from surrounding dialogues.
We find that the user often needs to refine further and enhance the clarity and engagement of the moment. 
They often incorporate new surrounding sentences, remove rambling sentences, or adjust the order of the sentences.

Figure \ref{fig:refine} illustrates our sentence-level interface on refinement.
We provide a list of the sentences from the selected moment, all selected by default.
Each sentence has the speaker’s headshot shown to help locate the speaker's transitions. 
The system also shows the relevant context before and after the moment. 
This context includes the neighboring surrounding sentences or disjoint sentences from a similar topic.

For example, if the selected moment is in the middle of a guest's speech, the system will automatically find the corresponding question from the host that leads to this speech, and vice versa. 
The guest might also elaborate on their statement in the following sentences. 
The system also incorporates the sentences right after the clip to help clarify the content. 
Sometimes, the leading question and the selected moment would be far apart. 
For example, if the guest has a long speech, the system allows the user to check the sentences right before the selected moment to help them understand the whole context. 
The user can check them by clicking the ``Show Between'' button. 

If the recommendations do not cover what the user wants, we provide a search bar to find the exact sentences they want.
The user can drag, drop, and rearrange sentences as desired.

It is important to keep the teaser concise within a reasonable duration. Thus, the system indicates the selected sentences' overall duration to give the user an idea of the current length.
Additionally, the system offers a ``remove filler words'' button to automatically detect and remove filler words like ``um'' and ``uh''.
This feature helps maintain a better flow and ensures conciseness.

After the user decides which sentences to incorporate and in what order, they can click ``Preview''. 
The system constructs these sentences into a refined clip for users to watch.

The first three steps pertain to content assembly. Next, we will discuss steps related to production. 
To support the production of all essential elements and give users control (D4), we provide tools to smooth transitions around cuts in the refine process, as well as adding music and branding to enhance the polish of the final product. To ensure users have full control over all details, they have all the Premiere Pro tools at their disposal. 

\subsection{Transitions}
After the user adjusts and reorders sentences on the refine tab, the video may have jump cuts.
Jump cuts are abrupt transitions from the same speaker that negatively affect the audience's viewing experience. 
To address this issue, creators often add transitions at the jump-cut locations \cite{berthouzoz2012tools}.

The system displays the sentences from the refinement step and detects if there are any jump cuts between them. 
For instance, if the system detects two disjointed sentences from the guest, it recommends adding transitions between them. 
If no jump cuts are detected, the system will suggest skipping this step and proceeding to the next.
We implemented two popular types of transitions to hide jump cuts: zoom-in/out effect and reaction shots.

A zoom-in/out effect at the jump-cut location can hide the jump cut and create a smoother transition. 
Clicking the ``add zoom in/out'' button triggers Premiere Pro to adjust the frame's scale before and after the jump cut. 
This reduces the visual misalignment brought by the jump cut. 
After this zoom effect is added, the system shows a magnifier icon to the scaled sentence to indicate the change.

Adding reaction shots is another way to hide the jump cut.  
Our system analyzes the visible person and the speaker's alignment to collect reaction shots. 
For example, if the system detects the guest is speaking while the host is visible, it indicates that the host is reacting to the guest's speech.
If the user clicks ``add reaction shot,'' the system will automatically find the nearest reaction shot and place it at the jump-cut location. 
A thumbnail of the reaction shot will be displayed to indicate the change in the extension.
Adding a reaction shot not only hides the jump cut but also enhances the dynamics of the clip since it shows the reactions, such as nodding and laughing, from other people in the conversation. 

The user can preview the revised clip by clicking ``Preview.'' 
They also have the option to remove the added effects. 

\subsection{Music}

\begin{figure}
\includegraphics[width=0.38\textwidth]{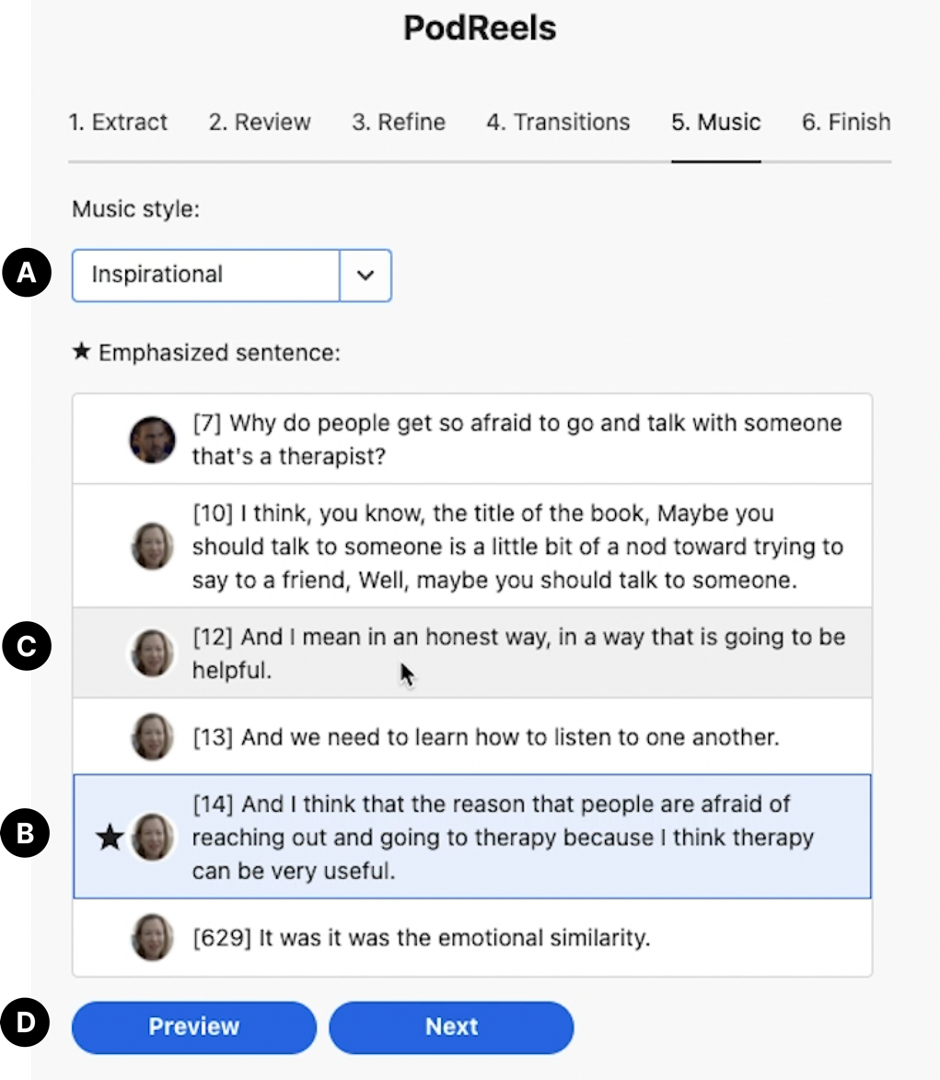}
\caption{The music step allows users to easily add music to their teaser to make it more engaging. (A) Users can pick a preset style from the list. (B) The system automatically identifies the peak (i.e., the emphasis point) in the transcript and highlights it with a star icon. (C) If users want to change the emphasis point, they can click on the new sentence. (D) Users can preview the teaser with music. Under the hood, PodReels has added and arranged the music to align with the emphasis point. For example, the system will insert the music’s intro/verse during non-emphasized sentences and insert a chorus on the emphasized sentence. 
}
\label{fig:music}
\Description{The music step allows users to easily add music to their teaser to make it more engaging. (A) Users can pick a preset style from the list. (B) The system automatically identifies the peak (i.e., the emphasis point) in the transcript and highlights it with a star icon. (C) If users want to change the emphasis point, they can click on the new sentence. (D) Users can preview the teaser with music. Under the hood, PodReels has added and arranged the music to align with the emphasis point. For example, the system will insert the music’s intro/verse during non-emphasized sentences and insert a chorus on the emphasized sentence.}
\end{figure}

The use of background music can significantly enhance the engagement of a teaser. 
Shown in Figure \ref{fig:music}, our system allows users to choose different styles of music that fit their teaser from a common set---inspirational, emotional, uplifting, and light-hearted.
One common technique to add background music is to align the peak of the music with the emphasis point of the speech \cite{rubin2013content}. 
This can help the audience sense the emotional change and highlight key moments. 
The system uses the LLM to detect the sentence to emphasize, which is highlighted with a star icon. 
If the user is unsatisfied with the recommendation, they can adjust the emphasized sentence by clicking on other sentences.

We then align the ``peak'' of the music to this emphasized sentence.
We manually labeled and split each piece of music into two parts: regular patterns and peaks. 
After the user decides on the emphasized sentence, we generate music that covers the entire length by repeating the regular pattern music pieces and adding the peak at the emphasis point's timestamp. 
This simple technique helps align the peak of the music with the emphasized sentence. 

Users can preview the clip with the added background music by clicking ``Preview.'' 
They also have the option to not use music by choosing ``None'' under the music style. 

\subsection{Finish}

Finally, PodReels assists the user in the production steps to create their final teaser. For this final step, PodReels leverages tools already available in Premiere Pro, like automatic framing for different aspect ratios, captions, and graphics. 

Since teasers are often published on social media platforms such as TikTok, Instagram Reels, and YouTube Shorts, PodReels offers vertical, square, and horizontal aspect ratios. To automatically crop the footage to center each speaker, PodReels uses the Premiere Pro built-in Auto Reframe feature\footnote{\url{https://helpx.adobe.com/premiere-pro/using/auto-reframe.html}}. 
Additionally, PodReels lets the user automatically add captions to the video through one of two styles: ``standard'' and ``rapid'', which are often used in social media reels.
Standard captions display five words per line and rapid captions display two words per line. 
The user can also leverage other present caption styles that come with Premiere Pro that allow changing the font style, size, and location.
Last, PodReels automatically adds branding graphics (e.g., the podcast channel's logo) through the ``Add logo'' button.

\section{Technical evaluation}
To evaluate the effectiveness of large language models (LLMs) in finding clips according to user criteria (duration, featured speakers, style, and content keywords), we measure GPT-4's performance for finding clips with a single parameter and with multiple parameters.
Note that the technical evaluation results pertain to GPT-4 (gpt-4-32k-0613), and the accuracy results may become outdated as LLMs continue to evolve. 
The primary goal of this study is to assess the efficiency of GPT-4 as an underlying model of PodReels. 
PodReels can use any LLM as the backend.

\subsection{Methodology}
For the evaluation dataset, we randomly selected four podcast episodes from YouTube's trending podcasts page, each corresponding to a different genre: society\&culture, sports, health\&fitness, and education, with an average duration of 45 minutes. 
We used the Extract (Section \ref{sec:extract}) and Review (Section \ref{sec:review}) tabs of the PodReels system to import episode videos, specify search parameters, and extract clips. 

For the single-parameter evaluation, we evaluated the four parameters separately. 
The parameter options included three durations (15s, 30s, 45s), three speaker selections (host-only, guest-only, both host and guest), four styles (informational, curiosity-arousing, funny, emotional), and four keywords (the first four of the recommended keywords). 
We queried the LLM for three clips per option.
Altogether, we collected 168 clips across the four episodes: 36 based on duration, 36 based on speaker selections, 48 based on styles, and 48 based on keywords.

For the multi-parameter evaluation, we randomly selected 30 possible combinations to ensure parameter diversity in our sampling. 
(Our dataset comprises 576 possible combinations across the aforementioned parameter options for four episodes.) 
We queried the LLM with each combination for one clip, yielding 30 total clips.

To evaluate the duration and speaker parameter results, one author watched the clips and recorded the duration and speakers, while another author performed a recheck. 
Since this is an objective judgment, the two authors reached a perfect agreement on the annotation. We consider an error within a five-second margin acceptable for duration accuracy. To evaluate the style and keyword parameter results, we recruited four annotators (two female and two male, average age 24.0) through a university emailing list. 
The annotators were video podcast enthusiasts who watched podcasts daily.
We asked the annotators to watch the clips and rate the relevance of the style or keyword.
For each clip, the annotators independently assigned scores on a 7-point scale. 
Two annotators rated the single-parameter set, while the other two rated the multi-parameter set. 
The inter-rater reliability was Cohen’s $\kappa$ = .61 and .63, respectively, showing substantial agreement between the annotators’ judgments.
We averaged the scores from the two annotators and considered five or higher to indicate a successful match.
The annotators were compensated \$25/hour.

\subsection{Single-parameter Results}
Overall, GPT-4 demonstrates a high capability for finding clips using a single parameter (see Table \ref{tab:annotation_results}), achieving an accuracy of 86.1\% for duration, 88.9\% for speakers, and 87.5\% for keywords. 
For duration, we considered clips exceeding the five-second margin as inaccurate. 
Across the clips, such inaccuracies did not surpass ten seconds. 
Regarding speakers, all errors occurred in the ``host\&guest'' condition. 
There were no errors when searching for clips containing host or guest only.
For keywords, annotators noted that they assigned low relevancy scores for some clips because, despite one or two sentences in the clip being relevant to the keyword, the main topic of the entire clip did not align with the searched keywords.

The accuracy for style-based parameters is slightly lower at 77.1\%, with specific performance varying depending on the style.
In our experiment, GPT-4 performs exceptionally well in identifying curiosity-arousing style clips, achieving perfect accuracy. 
Informational and emotional styles achieve 83.3\% accuracy. 
However, the accuracy for finding funny-style content is relatively low at 41.7\% (7 out of 12 clips fail).

The lower accuracy for the funny style was influenced by the lack of such content in one of the health\&fitness episodes.
This episode discusses veganism and compassion for animals in a sad tone and all three ``funny'' clips from this episode received a score of 1 (out of 7) from the annotators. 
One author watched the whole episode and verified that there were no obviously funny excerpts.
Three of the remaining four clips that were not identified as funny received an average score of 4 (out of 7). 
The annotators noted that, while these clips featured speakers sharing their experiences light-heartedly, they did not resonate as very humorous. 

Overall, the LLM performs well in conducting single-parameter searches, even over relatively abstractive attributes like style. 
We discuss future work of identifying impossible parameter options in Section \ref{future_work}. 

\begin{table}[t] 
\centering 
\scalebox{0.9}{
\begin{tabular}{| l | c | c | c | c |} 
\hline
& \textbf{Duration} & \textbf{Speakers} & \textbf{Style} & \textbf{Keyword} \\ %
\hline       
\textbf{Single-parameter} & 86.1\% & 88.9\% & 77.1\% & 87.5\% \\
\hline
\textbf{Multi-parameter} & 83.3\% & 83.3\% & 73.3\% & 80.0\% \\
\hline
\textbf{Difference} & -2.8\% & -5.6\% & -3.8\% & -7.5\% \\
\hline
\end{tabular}
}
\caption{Accuracy of parameters (duration, featured speakers, style, and keyword) in single-parameter and multi-parameter conditions.} 
\label{tab:annotation_results} 
\Description{This table compares accuracy between single and multi-parameter clips across four parameters: duration, speakers, style, and keyword. In the first row, the accuracy percentages for single-parameter clips are listed as 86.1\% for duration, 88.9\% for speakers, 77.1\% for style, and 87.5\% for keywords. The second row shows the accuracy for multi-parameter clips, with percentages at 83.3\% for both duration and speakers, 73.3\% for style, and 80.0\% for keywords. The third row lists the percentage difference in accuracy between single and multi-parameter clips, showing a decrease in accuracy for multi-parameter clips: -2.8\% for duration, -5.6\% for speakers, -3.8\% for style, and -7.5\% for keywords.}
\end{table}

\subsection{Multi-parameter Results}

Overall, 43.3\% of clips score high in all four parameters, 33.3\% in three parameters, and 23.3\% in two of them.
None of the clips score high in only one or zero parameters.
Multi-parameter search is hard and needs trade-offs between parameters.
Considering the challenge of optimizing for all parameters simultaneously, we regard high scores for three parameters or more success.  
The LLM performs well across 76.7\% of the evaluated clips in at least three dimensions. 
In the cases where two parameters fail, the clips often fail in the featured speakers (mostly host\&guest) along with style (mostly funny) or keyword, which aligns with what was observed in the single-parameter study.

When compared to single-parameter searches, the LLM's accuracy across each parameter declines to varying degrees in the multi-parameter search scenarios (see Table \ref{tab:annotation_results}). 
Specifically, the LLM achieves an accuracy of 83.3\% for both duration and speakers, marking decreases of 2.8\% and 5.6\%, respectively. 
Furthermore, accuracy for style reaches 73.3\%, a decline of 3.8\%, and for keywords, the accuracy is 80.0\%, representing a decrease of 7.5\%.
Note that only 6.7\% of clips fail in both style and keyword dimensions.
We can see that the LLM strikes a reasonable balance between the parameters, with none of them dropping significantly, giving users a good starting point for their teasers.

\section{User Study}

To understand how PodReels might assist users in creating video podcast teasers, we conducted a within-subjects study with ten participants (five experts and five novices in terms of teaser creation), comparing PodReels to a baseline.

\begin{figure}
\includegraphics[width=0.48\textwidth]{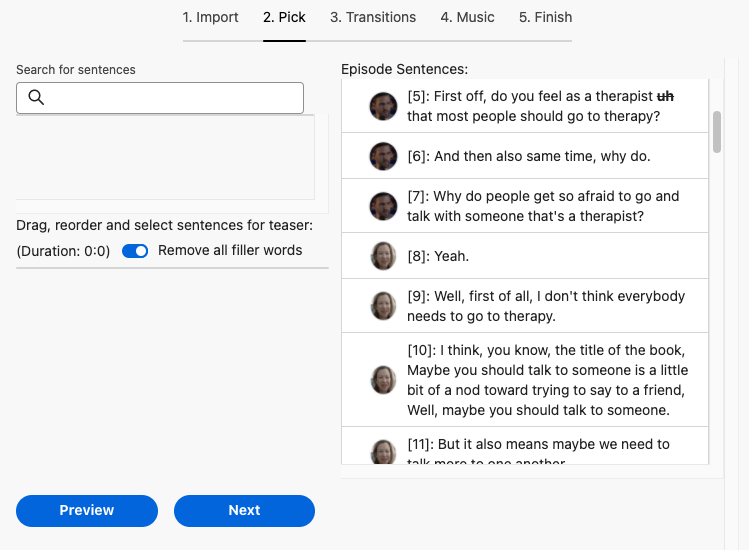}
\caption{Screenshot of the baseline system. The baseline has identical transitions, music, and finish features as of PodReels for production support. While for the content assembly part, it only provides basic transcript-based selection and preview functionalities. }
\label{fig:baseline}
\Description{Screenshot of the baseline. The baseline has identical transitions, music, and finish features as of PodReels for production support. While for the content assembly part, it only provides basic transcript-based selection and preview functionalities.}
\end{figure}

The baseline (see Figure \ref{fig:baseline}) shared a similar UI design with PodReels. 
It included identical features for the production phase---transitions, music, and finish. 
However, in the content assembly phase, the baseline did not offer the extract, review, and refine features available in PodReels. 
Instead, it provided basic transcript-based selection and preview functionalities. 
Users could browse the transcript on a sentence level and choose sentences for their teasers by dragging and dropping them. 
A search function was also available for locating specific sentences.
Users could still reorder or re-select sentences, check the current duration, remove filler words, and preview the outcome.

Each participant was asked to create two teasers for two podcast episodes, one for each, using each system. 
After finishing the teaser creation with each system, participants were asked to evaluate the system through questionnaires on outcome satisfaction, task load index, creativity support index, system usability, and usefulness. 
We also conducted semi-structured interviews to understand their experiences with each system. 
At the end of each study session, we conducted another interview to get feedback about their overall experience and system preferences.

\subsection{Hypotheses}
In the study, we investigate the following hypotheses:
\begin{enumerate}[nosep,leftmargin=*,label={\textit{H{\arabic*}}}]
    \item Compared to the baseline, PodReels significantly \textbf{decreases the time designers need to create a teaser (H1}). 
    \item Compared to the baseline, PodReels significantly \textbf{increases designers' satisfaction level (H2)} with the overall quality (H2a) of the teaser, including the quality of content assembly (H2b) and production (H2c).
    \item Compared to the baseline, PodReels significantly \textbf{lowers designers' work load (H3)} for mental demand (H3a), temporal demand (H3b), performance (H3c), effort (H3d) and frustration (H3e). 
    \item Compared to the baseline, PodReels provides significantly \textbf{better creativity support (H4)} in terms of exploration (H4a), engagement (H4b), effort/reward tradeoff (H4c), tool transparency (H4d) and expressiveness (H4e). 
    \item Compared to the baseline, PodReels has significantly \textbf{higher usability (H5)}. 
    \item Compared to the baseline, PodReels is significantly \textbf{more useful for designers (H6)}.
\end{enumerate}

\begin{table}[t]
\scalebox{0.85}{
    \begin{tabular}{| c | c | c | c | c | }
    \hline 
         \textbf{Participant} & \textbf{Podcast genre} &\textbf{Episode1 length} &\textbf{Episode2 length}\\ 
         \hline
         \textbf{E1}&  Education &  40min 31s& 44min 39s \\ 
         \hline 
          \textbf{E2}& Business &  57min 21s& 57min 48s  \\ 
         \hline 
        \textbf{E3}&  Health\&Fitness &  45min 44s & 48min 20s  \\ 
        \hline
         \textbf{E4}& Health\&Fitness & 1h 3min 16s& 1h 0min 54s   \\ 
         \hline 
         \textbf{E5}& Business & 51min 20s& 47min 48s   \\ 
         \hline 
         \textbf{N1}&  Sports  & 38min 3s& 33min 15s \\ 
         \hline 
         \textbf{N2}&  Leisure & 1h 2min 47s& 1h 2min 14s  \\ 
         \hline 
         \textbf{N3}&  Health\&Fitness  & 40min 31s& 40min 32s \\ 
         \hline 
         \textbf{N4}&  Society\&Culture& 43min 17s& 46min 31s \\ 
         \hline 
        \textbf{N5}&  Education&  59min 26s& 1h 1min 28s \\ 
         \hline 

    \end{tabular}
    }
    \caption{Overview of podcast episodes picked by participants, including podcast genre and length of each episode. The genre information was collected from the description of the according podcast channel on Apple Podcasts.}
    \Description{Table detailing podcast episodes selected by participants for an experiment. It lists participants coded E1 to E5 and N1 to N5, with their chosen podcast genres ranging from Education, Business, Health & Fitness, Sports, Leisure, to Society & Culture. The table includes the length of two episodes per participant, with times varying from 33 minutes 15 seconds to 1 hour 3 minutes 16 seconds.}
    \label{tab:user_picking_examples}
\end{table}

\subsection{Participants and Procedure}
We recruited ten participants through Upwork and interest-based social media forums. 
All participants owned podcast channels or were professional podcast editors. 
As PodReels is built within Premiere Pro, our screening criteria included a requirement for knowledge of basic Premiere Pro operations.
Among ten participants (average age=27.7, four female, six male), five are experts (E1-E5), and five are novices (N1-N5) in terms of their video podcast teaser creation experience.
On average, the five experts had two years of experience making video podcast teasers, and each had created at least 50 video podcast teasers. 
The novices had less experience and created fewer than five video podcast teasers before.

Before the study, we asked each participant to pick two video podcast episodes from one podcast channel they wanted to make a teaser for, with a length difference of under five minutes. 
The podcast episodes that each participant picked can be found in Table \ref{tab:user_picking_examples}. 
Participants had already watched their picked episodes before.

The study sessions were conducted remotely. 
Before each session, we asked participants to install Zoom and the latest version of Premiere Pro (v23.6 at the time of study). 
During the session, we first introduced the concepts of video podcast teasers and showed participants teaser examples. 
Then, each participant was asked to complete two teaser creation tasks using their episodes with the baseline and PodReels.
Experts and novices were randomly assigned to a condition order (either baseline first and then PodReels or PodReels first and then baseline) that was counterbalanced to prevent a learning effect. 
Before each task, we provided them with the extension installation package and the pre-processed ML data of the episode they selected. 
We used an example episode to guide participants through the system and familiarize them with the interface.
Then, participants were given a maximum of 20 minutes to complete each task. 
The entire process did not exceed 1.5 hours. 
Participants were compensated \$100/hour. 

\begin{table*}[t]
\centering
\begin{tabular}{llrrrrrrll}
\toprule
\multicolumn{1}{c}{\multirow{2}{*}{\textbf{Category}}}   &
\multicolumn{1}{c}{\multirow{2}{*}{\textbf{Factor}}}               & \multicolumn{2}{c}{PodReels}     & \multicolumn{2}{c}{Baseline}      & \multicolumn{2}{c}{Statistics} & \multicolumn{1}{c}{\multirow{2}{*}{\textbf{Hypotheses}}}\\
\cmidrule(lr){3-4}\cmidrule(lr){5-6}\cmidrule(lr){7-8}
           &                      & Mean    & SD    & Mean     & SD    &  $p$               & Sig. \\
\toprule

\multicolumn{1}{l}{\multirow{1}{*}{\textbf{Time}}}   
& Time used (in minutes) & 9.9 & 2.23 & 16.7 & 3.06 
& .001  & *** & \textit{H1} accepted\\
\midrule
\multicolumn{1}{l}{\multirow{3}{*}{\textbf{Satisfaction}}}    & Overall outcome  & 5.6 & .52 & 4.1 & .99  
& .004  & **  & \textit{H2a} accepted\\
& Content Assembly & 6.2  & .63 & 3.4 & 1.17 
& .003  & **  & \textit{H2b} accepted \\
& Production & 4.5 & 1.27 & 4.3 & .95 
& .283 & -  & \textit{H2c} rejected \\
\midrule
\multicolumn{1}{l}{\multirow{5}{*}{\textbf{Task load}}} & Mental demand & 2.4  & .52  & 5.4 & 1.17 
& .003 & ** & \textit{H3a} accepted \\
& Temporal demand & 2.7 & .67 & 4.5     & .85  
& .002 & **  & \textit{H3b} accepted     \\
& Performance  & 2.5  & .53  & 4.2  & .79  
& .004  & **  & \textit{H3c} accepted    \\
& Effort & 2.8  & .79  & 4.7 & .82   
& .002 & **  & \textit{H3d} accepted    \\
& Frustration & 2.0 & .67 & 4.3 & .95   
& .003  & ** & \textit{H3e} accepted   \\
\midrule
\multicolumn{1}{l}{\multirow{5}{*}{\textbf{Creativity support}}} 
& Exploration  & 6.4  & .52  & 2.2 & .63 
& .002  & **  & \textit{H4a} accepted   \\
& Engagement  & 6.1 & .74 & 3.5 & .53  
& .003 & **  & \textit{H4b} accepted     \\
& Effort/reward tradeoff & 5.5 & .85 & 3.3 & .67  
& .006  & **   & \textit{H4c} accepted    \\
& Tool transparency  & 3.8 & .92  & 3.2 & .63   
& .024 & *  & \textit{H4d} accepted    \\
& Expressiveness  & 5.1 & .88  & 2.9     & .74   
& .004 & ** & \textit{H4e} accepted   \\
\midrule
\textbf{Usability}  &  & 79.75 & 5.46 & 77.25 & 3.99 
& .138  & -  & \textit{H5} rejected       \\
\midrule
\textbf{Usefulness} &  & 6.4 & .52  & 4.2 & .79  
& .002 & ** & \textit{H6} accepted \\
\bottomrule
\end{tabular}
\caption{The statistical test results comparing PodReels with the baseline, where the p-values ($-$: $p>.100$, $+$: $.050<p<.100$, $*$: $p<.050$, $**$: $p<.010$, $***$: $p<.001$) are reported. }
\label{tab:study_result}
\Description{This table presents the results of a comparative study between PodReels and a baseline system across several categories, including time, satisfaction, task load, creativity support, usability, and usefulness. Data points such as the mean and standard deviation (SD) for each category are provided for both PodReels and the baseline. The statistical significance is denoted by p-values and significance levels indicated as -, +, *, **, and ***. Significant findings show PodReels generally performs better in terms of time used, satisfaction, task load, and creativity support, with many hypotheses accepted in those categories. However, in the usability category, the hypothesis is rejected, with no significant difference found. The usefulness category shows significant improvement, with the hypothesis accepted.}
\end{table*}

\subsection{Results Analysis}
We collected participants' ratings on a 1-7 point scale through questionnaires (see Table \ref{tab:study_result} for results).
Participants rated their outcome satisfaction, task load, creativity support, usability, and usefulness in using PodReels and the baseline. 
We recorded the time each participant spent on each task.
During the interviews, we asked them follow-up questions to understand the reasons behind their scores. 
We applied the thematic analysis method~\cite{braun2006using} to analyze the interview transcripts. 
The initial coding was performed by the first author, and the research team collaboratively reviewed the coding results to determine the themes.
We report the key findings in this section.

\subsubsection{Time needed to create a teaser}

On average, participants spent 9.9 minutes (SD=2.23) creating a video podcast teaser using PodReels, roughly half the time they spent with the baseline (mean=16.7, SD=3.06).
Participants generally appreciated the sentence-level picking and previewing features in the baseline. 
However, they all spent significant time rewinding and reviewing the podcast video, manually targeting interesting clips.
All participants agreed that the \emph{extract} feature in PodReels is a great timesaver, preventing them from repeatedly checking the hour-long podcast video to locate potential clips. 
As N3 pointed out, \textit{``[Without PodReels..] you just don't even know what [clip] to pick. But this, you know, presenting three options at a time makes it easier to narrow down your choices. And the options are good. It speeds up the process a lot.''} 
The baseline offers keyword searches but is often insufficient for creators to target interesting clips.
They often need to consider other dimensions like duration, style, and featured speakers; PodReels facilitates search that integrates these dimensions, helping users more efficiently find the clips they want.

Additionally, participants appreciated how the \emph{refine} page provided a window for them to understand the context and find supplementary statements. 
E1 stated, \textit{``The before and after [context] sentences on the refine page were very helpful. Sometimes, it's difficult to get the clip's content as it ends a bit too early. However, these surrounding sentences made it very feasible for me to understand and provide context for the audience.''}

\subsubsection{Perceived quality of outcome}
Teasers created with PodReels had a significantly higher perceived overall quality (mean=5.6, SD=.52) than with the baseline (mean=4.1, SD=.99), based on participants' self-reported results.
E2 liked how PodReels sped up their process while preserving the outcome quality---\textit{``I definitely want to incorporate this [PodReels] into my future workflow. It can help me put out quality content more frequently for sure.''}
Creators also enjoyed how they had control over every single sentence in the outcome.
E4 shared \textit{``Everyone [on social media] has such a short attention span. So every single moment within the reel matters. And this gives you the ability to really have control over what is being shared in the reel and then in a very easy way, a very smooth way.''}

Next, we further break down the quality perception into content assembly and production.
Participants perceived a significantly higher content assembly quality in their teasers created with PodReels (mean=6.2, SD=.63) compared to the baseline (mean=3.4, SD=1.17). 
As N4 mentioned, the \textit{review} page provided clip candidates of great quality. 
\textit{``The clips pulled out by the system made a lot of sense to me. I definitely saw quite a few options which were already very engaging at a reasonable length; the liveliness indicator is interesting as well."} shared by N4. 
Additionally, participants noted that the \textit{refine} page helped them decide how much information to reveal in the teaser. 
As E3 summarized, \textit{``There's just like that fine line where you want it to cut off to drive the viewer's curiosity, but you also don't want them to be like, wait, what? That was it? The refine page really helped me to strike that balance.''} 

There was no significant improvement in the production quality of the outcome when creating with PodReels. This is expected because the baseline had the same production phase features as PodReels.  
We discuss participants' feedback on production features in Section \ref{sec:suggestions}.

\subsubsection{Task load}
In the NASA TLX dimensions~\cite{hart2006nasa}, creating with PodReels was significantly less demanding in terms of mental demand (p=.003, Z=-2.78), temporal demand (p=.002, Z=-2.82), performance (p=.004, Z=-2.69), effort (p=.002, Z=-2.86), and frustration (p=.003, Z=-2.78).
PodReels greatly eased the burden of searching for compelling clips. 
E2 pointed out PodReels reduced mental effort: \textit{``Even though I had a general idea of what kind of clips I wanted, manually looking for them was very tiring. I had to rewatch the episode multiple times and be very focused. It seems impossible to find a perfect clip that checks all my boxes at once. So I was really relieved when the system was able to conduct the search for me.''}
E3 and E4 also shared that being able to quickly see the clip options reduced mental load and frustration:
E4 said \textit{``So I won't be stuck in searching for clips that might not exist''}.
E3 said PodReels was much less mentally demanding than other AI tools they had used---\textit{``The other program I used before just spun out so many different clip options. It was very overwhelming. Seeing three at a time [in PodReels] made it much easier for me to make the decision.''}

\subsubsection{Creativity support index}
PodReels  provided significantly better creativity support \cite{csi} in terms of exploration (p=.002, Z=2.82), engagement (p=.003, Z=2.8), effort/reward tradeoff (p=.006, Z=2.49), tool transparency (p=.024, Z=1.98), and expressiveness (p=.004, Z=2.63).
Participants appreciated PodReels allowing them to explore diverse options and customize the clip to their liking.
For example, E3 shared \textit{``The AI software I tried before doesn't give me that freedom [of customizing]; They are like. This is what it is. Take it or leave it. Whereas this [PodReels] gives you the ability to explore possible options, and mix and match without making it overly complicated.''}

In addition, the automation of some tedious editing steps helped participants boost their engagement by only having to focus on high-level decisions and operations. 
For example, N2 shared that the preview feature was very useful to them, stating, \textit{``In a second, the sentences I picked were being constructed together and I could see how they worked. This was really smooth and cool. So now I just needed to focus on thinking which combination of the sentences was good instead of worrying about tracking their timestamps and doing all the locating, trimming, and sequencing [tasks].''}

\subsubsection{Usefulness of the system}
According to participant ratings, PodReels  (mean=6.4, SD=.52) was significantly more useful than the baseline (mean=4.2, SD=.79). 
The workflow of extract, review, and refine in PodReels provided participants with a flexible way to assemble clips of high quality.
Nine out of ten participants mentioned their likes over the \textit{refine} page, making it easy for them to do the necessary refinement.
In addition, participants appreciated that they could work in a non-linear way to explore the clip styles and content.
As N1 shared, \textit{``I was initially thinking the keyword of relationship and informational style would be a good combination, but seeing the options helps me decide if we chose emotional, it would have been better.''}

\subsubsection{Usability of the system}
To measure the usability of the systems, We adopted the standard System Usability Scale (SUS) questionnaire \cite{brooke2013sus}. 
The SUS rating of the PodReels is 79.75 (A- level), while the baseline rates 77.25 (B+ level) \cite{bangor2009determining}, both at an acceptable level with no statistically significant difference.
This result is expected, as both systems support transcript-based editing and share a similar UI. 
Participants gave positive comments for both systems in terms of their ``streamlined workflow'' (E1, E3, E4, E5, N2, N3), ``simple and intuitive interface'' (E1, E2, E4, N1, N2, N3), and ``beginner-friendly design'' (N1, N3, N4, N5).
We discuss the limitations of the system development in terms of usefulness and usability in Section \ref{sec:develop_plugin}.

\subsubsection{Additional feedback}
\label{sec:suggestions}

All participants mentioned that the suggested moment clips were generally of good quality. 
However, sometimes certain clip options might ``not match the style keyword as well as expected'' (E4) or ``could be more accurate in terms of length'' (E2, N3). 
Generally, all participants were able to find at least one suitable
candidate within one or two rounds of search.

All participants agreed that PodReels provided them with basic necessary production support. 
However, three expert creators (E2, E3, E4) mentioned they would want more support in terms of music, as they saw music as a very important way to add engagement to the social media reels. 
They stated that they would adjust the music in Premiere Pro before publishing the teaser.
Currently, in PodReels, we provide one music track for each of the four styles and alignment support. 
E4 commented that \textit{``The alignment feature is nice, but you need more music. Like you’ll need more of each kind, so you can maybe have a dropdown from each.''}
The other two expert creators (E1 and E5) mentioned that they would adjust the caption styles in Premiere Pro to better align with the brand style of their podcast.

Participants also suggested we integrate more interactions into PodReels. 
For example, currently, users can begin playing and previewing clips by clicking the clip’s thumbnail in PodReels. 
However, if they wish to pause, they need to use the player control button in the video player. 
We discuss this further in Section \ref{sec:develop_plugin}.

\section{Discussion}
\subsection{Developing AI-powered Extensions to Professional Tools}
\label{sec:develop_plugin}
PodReels is an AI-powered extension to Premiere Pro, a professional video editing tool. 
PodReels supports podcasters in creating teasers by streamlining the content assembly and production steps.
Our study shows that PodReels succeeds in extracting clips that satisfy desired criteria with nearly 80\% accuracy, decreases users' mental demand by 41\%, and the time spent by 56\%.

Extracting clips that satisfy multiple desired parameters can be a daunting task for human creators alone. 
In our extension, we explored the use of AI to support this process. 
Large language models (LLMs) turn out to perform well in such multi-parameter semantic searches, even for relatively abstract goals like style matching. 
Although we did not specifically fine-tune the LLM for this task, it appears to possess an innate ability to balance multiple design goals, making reasonable trade-offs when not all of them can be fully met. 
Making trade-offs is a hallmark of many creative tasks---often, there is no perfect solution, just solutions with different trade-offs. 
For users who might struggle with these challenging design problems, incorporating LLMs into their creative process is an interesting direction to explore.

There are both pros and cons to our choice of building PodReels as an AI extension, instead of a standalone tool.
One of the biggest advantages is that the AI extension can offer users ``automatic AI magic,'' while still allowing them the option to revert to the host application for more control. 
For example, if users are not satisfied with the transition suggested by our extension, they can return to the host app and insert one of their choice.
Additionally, for low-level interactions that our extension currently does not support, such as adjusting the volume of background music, users can easily do that within the host app.
Another pro is that the extension can utilize the host app's infrastructure to support professional-level creations. 
For example, unlike existing standalone web applications with one- or two-track video players ~\cite{disco,truong2021automatic,berthouzoz2012tools}, our extension can access the robust multi-track video player in Premiere Pro, which is essential for creators aiming to produce teasers of professional audio and visual quality.

However, one of the cons is that we are heavily limited by what is exposed by the host application. 
We must adhere strictly to the data type and format exposed by Premiere Pro's API. 
For example, accessing the raw video frames in real-time is challenging due to the lack of available API.
By utilizing screenshots and audio captures, we might be able to leverage external APIs to acquire the necessary data.
Also, Premiere Pro does not allow programmatic changes to the caption style and control over the program monitor playback.
These constraints limit the features and interactions that can be implemented within the extension.

\subsection{Limitations and Future Work}
\label{future_work}

We built PodReels for video podcasts and it does not work on all video genres. 
First and foremost, we have only researched speech-heavy long-form videos with little visual changes. 
Additional investigation is required to generalize to videos where speech is less dominant and visual elements play a major role.  
We only focused on English podcasts, since parts of our ML pipeline (e.g., filler word detection and speaker diarization) only work for English. 
In the future, as machine learning algorithms work robustly for other languages, we hope to investigate how well our findings generalize to non-English podcasts.

As suggested by creators in Section~\ref{sec:creator-single-moment}, we built our workflow around single-moment teasers. 
Although all our participants make single-moment teasers, our sample size is small, and we do see multi-moment teasers in the wild, such as E4-E6 in Table \ref{tab:exsiting_examples}. 
Moreover, not all podcast teasers are based on pull-out moments---some podcasters would summarize the teasing moments, write them down, and record a separate video as teasers. 
This is analogous to the concept of abstractive summary (non-pull-out based teasers) and extractive summary (pull-out based teasers)~\cite{dalal2013survey}. 
Each direction has its own pros and cons---for video editing, the conventional methods are extraction-based since it is much easier to trim from existing video than generate them from a vacuum. 
However, with the recent advance of AI, it is possible that abstractive summaries combined with AI-generated videos could become a viable workflow.

We recognize the potential ethical issues that can arise from our work. 
In the formative study, we emphasized the importance of avoiding ``clickbaity'' teasers, which could erode the brand's reputation in the long term. 
While making minor adjustments to the original content for enhanced clarity and flow is acceptable, remixing video content to fit an unintended narrative can lead to misinformation. 
This can occur when the original context is obscured or altered, potentially spreading false or misleading information. 
To mitigate this, in the future, we could encourage users to compare their teaser draft with the original podcast experts.
We could flag the significant differences and nudge users to resolve them before publishing the teaser.

In the technical evaluation, we observe instances where clips of certain styles are absent in the episode video. 
As the reasoning ability of LLMs advances, we see their potential to assist users in identifying mismatches between desired parameters and available content. 
In our initial experiments on the episode with a sad tone, the LLM assigned ratings of 0 or 1 (out of 10) in terms of style for the ``funny'' clips it extracted. 
Through a more systematic investigation, we can explore ways to enhance our support in evaluating the alignment between the results and the desired design goals.

As reported in our formative study, high-quality production is a key ingredient for a good video podcast teaser. 
In addition to sentence selection and content assembly, we made an implementation choice to focus on tablestake production features like auto reframe, caption style, and static logo insertion, since these are highly requested features by creators. 
However, more advanced features can further boost production, including personalized image/video B-roll recommendations, more sophisticated text effect animation, and more.
In \textit{transitions} stage, we did not access the quality of the actual reaction---the system does not know if the detected shot is a positive/negative/neutral reaction. 
Thus, the reaction shot may not be quite right, and we rely on human input to accept or remove the reaction shot suggestion.
We look forward to further investigating these more advanced productions and designing the corresponding interactions that allow users to realize their creative vision.

\section{Conclusion}
We introduce PodReels, a human-AI co-creation system that assists video podcasters in creating teasers. 
We first investigate what makes a good teaser by combining insights from interviews with podcast listeners and creators. 
We identify a shared workflow among creators.
PodReels incorporates these insights to streamline the content assembly and production process of creating the video podcast teasers. 
Our study demonstrates that PodReels greatly reduces the time and mental demand required for creating teasers, resulting in a more enjoyable design process and more satisfying results.

\bibliographystyle{ACM-Reference-Format}
\bibliography{sample-base}

%TC:ignore
\appendix
\section{Mapping of Technical Components to PodReels Features}
\setcounter{table}{0}
\renewcommand{\thetable}{A\arabic{table}}

Table \ref{tab:tech_pipeline} shows the mapping of technical components to the features of PodReels.

\begin{table*}[t]
    \footnotesize
    \centering
    \renewcommand{\arraystretch}{1.1}
    \begin{tabular}{|l|l|c|c|c|c|c|c|c|c|}
    \hline
        ~ & ~ & \multicolumn{2}{c|}{Semantics} & \multicolumn{2}{c|}{Audio} & Visuals & \multicolumn{3}{c|}{Others} \\ \cline{3-10}
        ~ & ~ & \shortstack[c]{Transcription with \\ word-level timing} & \shortstack[c]{Filler word \\ detection} & \shortstack[c]{Active speaker\\diarization} & \shortstack[c]{Audio event \\ detection} & \shortstack[c]{Frame-level face \\ /body detection} & \shortstack[c]{GPT-4 \\ -32k} & \shortstack[c]{Google \\ Trends} & \shortstack[c]{Premiere \\ Pro} \\ \hline
        \multirow{2}{*}{1-Extract} & Keyword suggestion & X & ~ & ~ & ~ & ~ & X & X & ~ \\ \cline{2-10}
        ~ & Clip suggestion & X & ~ & X & ~ & ~ & X & ~ & ~ \\ \hline
        \multirow{3}{*}{2-Review} & Tagline generation & X & ~ & ~ & ~ & ~ & X & ~ & ~ \\ \cline{2-10}
        ~ & Liveliness score & ~ & ~ & ~ & X & ~ & ~ & ~ & ~ \\ \cline{2-10}
        ~ & Speakers featured & ~ & ~ & X & ~ & ~ & ~ & ~ & ~ \\ \hline
        \multirow{2}{*}{3-Refine} & Sentence suggestion & X & ~ & X & ~ & ~ & ~ & ~ & ~ \\ \cline{2-10}
        ~ & Filler word removal & X & X & ~ & ~ & ~ & ~ & ~ & ~ \\ \hline
        4-Transitions & Reaction shot suggestion & ~ & ~ & X & ~ & X & ~ & ~ & ~ \\ \hline
        5-Music & Emphasis suggestion & X & ~ & ~ & ~ & ~ & X & ~ & ~ \\ \hline
        \multirow{2}{*}{6-Finish} & Caption generation & X & ~ & ~ & ~ & ~ & ~ & ~ & ~ \\ \cline{2-10}
        ~ & Auto reframe & ~ & ~ & ~ & ~ & ~ & ~ & ~ & X \\ \hline
    \end{tabular}
    \caption{Mapping of technical components to the features of PodReels.}
    \label{tab:tech_pipeline}
    \Description{This table titled "Mapping of technical components to the features of PodReels steps" is organized into a structured format to display how various technical components are utilized across different steps in a media editing process. The table is divided into eight columns and multiple rows, categorized under phases labeled from "1-Extract" to "6-Finish." Each phase has sub-features like "Keyword suggestion," "Clip suggestion," and "Caption generation." Technical aspects such as semantic processes (including transcription with word-level timing and filler word detection), audio processes (like active speaker diarization and audio event detection), visual processes (like frame-level face/body detection), along with tools like GPT-4, Google Trends, and Premiere Pro, are mapped across these features to show their application in each step. Checkmarks (X) indicate the use of a technical component in a specific feature. The table serves as an overview of how different technologies integrate into the PodReels editing workflow, providing a clear visual representation of the technical pipeline.}
\end{table*}

\section{GPT-4 prompts used in PodReels}
\label{appendix:prompts}
\subsection{Extract}
\subsubsection{Extract episode keywords}
``Provide six main topic keywords for the following transcript. 
Transcript:
\textit{sentenceIndex: sentenceText}''

\subsubsection{Extract clips}
``This is the transcript of a podcast episode.
Transcript:
\textit{sentenceIndex [sentenceDuration]: (speakerID) sentenceText...}.
Select consecutive sentences to create a clip that must be around \textit{pickedLength} seconds long.
The clip should only include the following speakers: \textit{pickedSpeakers}.
The clip should be \textit{pickedStyle}.
The clip should contain the keywords of \textit{inputKeyWords}.
The transcript is given as a list of sentences with ID and duration in seconds. Only return the sentence IDs to form the clip. Do not include full sentences in your reply. Must return three distinct and non-over-lapping options of such clips. Use the following format: [a, b, c], [m, n, q], [x, y, z].''

\subsection{Review}
\subsubsection{Generate clip taglines}
``Come up with a catchy and short tagline for each of the clips. \textit{clipsContentString}. The tagline should be less than ten words.''

\subsection{Music}
\subsubsection{Find the sentence to emphasis}
``This is the transcript of a podcast clip.
Transcript:
\textit{sentenceIndex: sentenceText…}
Select one single sentence as the emphasis point in the clip. 
The transcript is given as a list of sentences with IDs. Only return one sentence ID you think should be emphasized. Must not include full sentences in your reply. ''

%TC:endignore 

\end{document}